\documentclass[a4paper,fleqn]{cas-sc}

\usepackage[authoryear,longnamesfirst,numbers]{natbib}
\usepackage{amsmath,amsfonts}
\usepackage{url}
\usepackage{algpseudocode}
\usepackage{algorithm}
\usepackage{xcolor}
\usepackage{booktabs}
\usepackage{braket}
\usepackage{cleveref}
\crefname{equation}{}{}
\Crefname{equation}{Equation}{Equations}
\crefname{figure}{Fig.}{Figs.}
\crefname{table}{Table}{Tables}
\crefname{section}{Section}{Sections}
\crefname{algorithm}{Algorithm}{Algorithms}

\begin{document}

\shorttitle{GQC and AQC for PF}    
\shortauthors{Z. Kaseb et al.}  

\title [mode = title]{Performance Comparison of Gate-Based and Adiabatic Quantum Computing for AC Power Flow Problem}  

\author[1]{Zeynab Kaseb}[orcid=0000-0002-5142-290]

\cormark[1]
\ead{Z.Kaseb@tudelft.nl}

\author[2]{Matthias M\"oller}
\author[1]{Peter Palensky}
\author[1]{Pedro P. Vergara}

\affiliation[1]{organization={Electrical Sustainable Energy, Delft University of Technology, The Netherlands}}
\affiliation[2]{organization={Applied Mathematics, Delft University of Technology, The Netherlands}}
\cortext[1]{Corresponding author}

\begin{abstract}
We present the first direct comparison between gate-based quantum computing (GQC) and adiabatic quantum computing (AQC) paradigms for solving the AC power flow (PF) equations. The PF problem is reformulated as a combinatorial optimization problem. For the GQC approach, the Quantum Approximate Optimization Algorithm (QAOA) is employed, while for the AQC approach, the problem is formulated as an Ising model. Numerical experiments on a 4-bus test system evaluate solution accuracy and computational performance. Results obtained using QAOA are benchmarked against those produced by D-Wave's Advantage\texttrademark\ system and Fujitsu's latest-generation Digital Annealer, implemented through the Quantum-Inspired Integrated Optimization (QIIO) software. The findings provide quantitative insights into the performance trade-offs, scalability, and practical viability of GQC and AQC paradigms for PF analysis, highlighting the potential of quantum optimization algorithms to address the computational challenges associated with the operation of modern electricity grids in the fault-tolerant era.
\end{abstract}

\begin{highlights}
\item First direct comparison of gate-based and adiabatic quantum computing for PF analysis.
\item Formulation of the AC PF problem as a combinatorial optimization problem.
\item Implementation of QAOA on gate-based simulators and annealers on Ising machines.
\item Evaluation of solutions and computational performance on a standard test system.
\item Insights into consistency of quantum approaches with classical PF solvers.
\end{highlights}


\maketitle

\section{Introduction}\label{}

Power flow (PF) analysis is a foundational task in electricity grids, used to compute the complex voltages at all buses given specified loads, generation, and grid topology. These voltages determine the nodal active and reactive power as well as the current flows on each line, thereby underpinning system operation and planning~\cite{Saadat1999,Xiao2024}. In alternating-current (AC) grids, PF equations are governed by Kirchhoff's laws and lead to a set of nonlinear, nonconvex equations. In practice, since the AC PF equations cannot be solved analytically, practitioners rely on iterative numerical methods, e.g., Gauss-Seidel (GS) or Newton-Raphson (NR) to find steady-state voltage solutions~\cite{Arrillaga1998}. While these classical solvers usually succeed, they can fail in large-scale or ill-conditioned cases. For example, GS depends heavily on the initial guess and often diverges under certain operating conditions, whereas NR may fail to converge if the Jacobian matrix becomes singular. NR is also computationally intensive and can perform poorly under heavy loading or high renewable penetration~\cite{Mokryani2016}. In modern electricity grids with many distributed energy resources, such convergence failures can undermine reliability and lead to inaccurate solutions. Thus, there is an increasing need for algorithms for the PF problem that are both computationally efficient and numerically robust to account for the challenges of modern electricity grids~\cite{Tripathy1982,Tostado2021}.

To address the aforementioned challenge, a fundamentally different approach is to reformulate the PF problem as a combinatorial optimization problem by discretizing bus complex voltages using spin/binary decision variables, thus transforming the AC PF equations into an Ising model and/or quadratic unconstrained binary optimization (QUBO) representation, e.g.,~\cite{kaseb2025aqpf2}. This combinatorial reformulation allows for the use of optimization solvers, but it also renders the problem strongly NP-hard~\cite{Bienstock2019}. As the electricity grid grows, the number of decision variables increases rapidly, making exhaustive search intractable. In addition, combinatorial reformulation introduces a higher-order polynomial that must be quadratized, further increasing the complexity~\cite{kaseb2024power}. Thus, while combinatorial PF reformulation offers a new perspective, it poses a large-scale, NP-hard optimization problem with rapidly growing dimensions.

Quantum computing has recently emerged as a promising paradigm for tackling combinatorial optimization problems, among others, e.g.,~\cite{Ganesan2021,Ali2025}. By harnessing quantum superposition and entanglement, quantum algorithms can potentially explore exponentially large solution spaces more efficiently than classical approaches, e.g., \cite{Au-Yeung2023}. In other words, as the problem size increases, classical solvers tend to plateau, while quantum hardware, if idealized, could maintain an advantage. However, current devices are in practice limited~\cite{PareekDemystifyingAdvantage}. Contemporary quantum devices are divided into two main paradigms: gate-based (circuit model) quantum computing (GQC) and adiabatic (quantum annealing) quantum computing (AQC). GQC hardware uses sequences of quantum gates on qubits in discrete time steps. They are very flexible in algorithm design and, in theory, can offer speedups for certain problems. For example, the Harrow-Hassidim-Lloyd (HHL) algorithm can solve linear equations in subexponential time, provided enough qubits are available. However, GQC algorithms tend to require many qubits and deep circuits even for modest problem sizes, which makes them highly susceptible to decoherence and gate errors on present noisy intermediate-scale quantum (NISQ) hardware. 

In practice, the limited qubit counts and noise levels mean that most GQC experiments are performed on small test problems or on high-fidelity simulators rather than on real quantum hardware~\cite{Imre2014,kaseb2024quantum}. On the other hand, AQC, exemplified by quantum annealers, evolves a quantum system continuously from an easy-to-prepare ground state to the ground state of a problem Hamiltonian. Adiabatic devices, so-called Ising machines, directly implement energy minimization on spin/binary decision variables. They are generally more noise-tolerant because the computation is analog and not gate-based~\cite{Kaseb2025SolvingComputing}. For example, D-Wave's Advantage\texttrademark\ system (QA)\footnote{\url{www.dwavequantum.com}} contains on the order of 5,000 superconducting qubits with sparse connectivity. Fujitsu's latest generation Digital Annealer (DA), i.e., Quantum-Inspired Integrated Optimization software (QIIO)\footnote{\url{en-portal.research.global.fujitsu.com/kozuchi}}, can emulate annealing on up to 100,000 fully-connected binary variables at room temperature~\cite{kaseb2024power}. These Ising machines specialize in solving NP-hard quadratic optimization problems. Nevertheless, they do not guarantee a perfect solution, but often find high-quality minima of the associated problem Hamiltonian~\cite{kaseb2025aqpf2}.

\begin{figure}[t]
\centering
\includegraphics[width=4in]{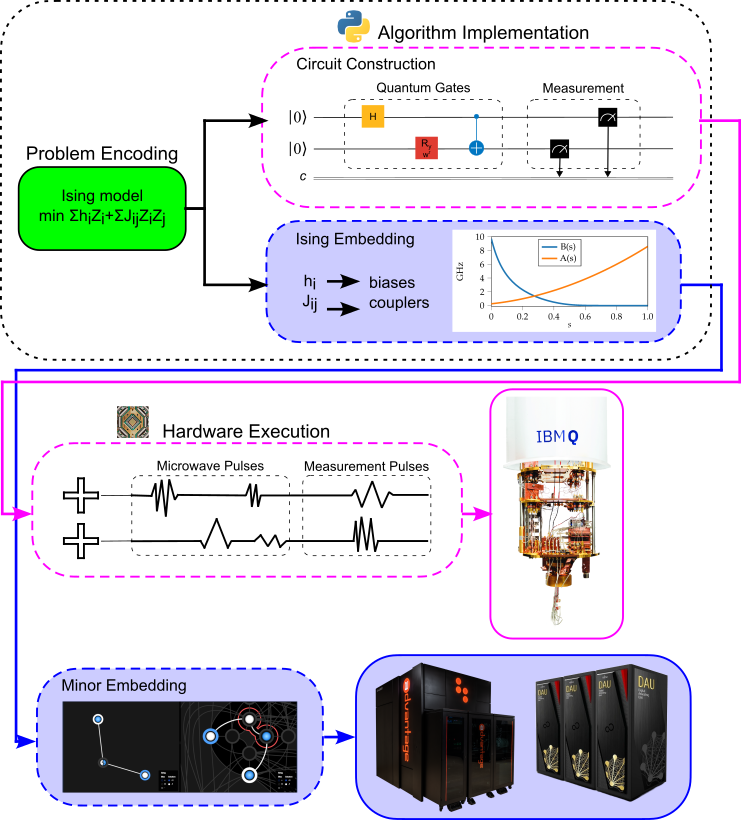}
\caption{Workflow for solving combinatorial optimization problems using GQC and AQC devices. The process consists of three stages: problem encoding, algorithm implementation, and hardware execution. In the encoding stage, the target optimization problem is formulated as an Ising model. During algorithm implementation, the problem is mapped to quantum circuits for GQC or embedded onto the hardware graph for AQC. In the execution stage, microwave pulse sequences are applied to gate-based quantum processors, e.g., IBM hardware, while minor-embedded Ising models are realized through programmable couplers on devices such as QA and DA.}
\label{fig:scope}
\end{figure}

Both paradigms have been applied to combinatorial optimization problems in the literature, e.g.,~\cite{Dupont2023,Armas2024}. A popular GQC approach is the Quantum Approximate Optimization Algorithm (QAOA), a hybrid quantum-classical variational algorithm that encodes the cost of a combinatorial optimization problem into a parameterized quantum circuit (PQC)~\cite{Pelofske2023,Farhi2022}. In theory, QAOA can yield better approximation ratios than classical heuristics for problems, such as Max-Cut or graph partitioning~\cite{Jing2023}. Nevertheless, to date, QAOA has only been implemented on small problem sizes, usually via simulation. Though quantum/digital annealers have shown promise on optimization benchmarks. For instance, a recent study comparing QAOA and AQC found that the analog annealer outperformed the GQC protocol on available machines~\cite{Pelofske2024}. \cref{fig:scope} illustrates the process of solving combinatorial optimization problems using GQC and AQC devices. The workflow comprises problem encoding, algorithm implementation, and hardware execution. Problem encoding refers to formulating the target optimization problem as an Ising model. Algorithm implementation includes quantum circuit design for GQC, and embedding for AQC. Hardware execution involves applying microwave pulse sequences to GQC hardware and realizing minor embedded via programmable couplers on AQC hardware. The figure illustrates IBM hardware for the former, and QA and DA for the latter.

In our prior study~\cite{kaseb2024power}, we introduced a combinatorial PF analysis by discretizing the AC PF equations into an Ising model and a QUBO representation and solved it using Ising machines. Numerical experiments on small test systems demonstrated that QA and QIIO can indeed recover accurate PF solutions and handle ill-conditioned cases. However, to date, there has been no GQC implementation or comparative study of the proposed combinatorial PF problem. In particular, it is unclear how QAOA (on a GQC hardware or simulator) would perform relative to annealing-based hardware for the PF problem. To address this gap, the present study employs both paradigms to solve AC PF equations. We implement QAOA using PennyLane's \texttt{lightning.qubit} statevector simulator\footnote{\url{docs.pennylane.ai/projects/lightning/en/stable/lightning_qubit/device.html}} for a standard 4-bus test system~\cite{Grainger1994}, and run the same problem on two Ising machines, i.e., QA and QIIO. We then compare the three solvers in terms of solution accuracy and computational time. The main contributions of this paper are: (i) providing the first implementation of the combinatorial PF problem using QAOA, and (ii) presenting a comprehensive comparison between GQC and AQC for the AC PF problem in the NISQ era.

\section{Combinatorial Power Flow Analysis}\label{}

Power flow (PF) analysis aims to determine the complex voltages within an electricity network and the associated power injections such that the system satisfies the steady-state power balance equations:
\begin{subequations} \label{eq:pf-balance}
    \begin{align}
        P_i & = P_i^{\mathrm{G}} - P_i^{\mathrm{D}}, \quad \forall i \in \{1,\dots,N\}, \label{eq:active-balance} \\
        Q_i & = Q_i^{\mathrm{G}} - Q_i^{\mathrm{D}}, \quad \forall i \in \{1,\dots,N\}. \label{eq:reactive-balance}
    \end{align}
\end{subequations}

Here, $N$ is the number of buses; $P_i$ and $Q_i$ are respectively the net active and reactive power at bus $i$; $P_i^{\mathrm{G}}$ and $Q_i^{\mathrm{G}}$ are respectively the generated active and reactive power at bus $i$; $P_i^{\mathrm{D}}$ and $Q_i^{\mathrm{D}}$ are respectively the consumed active and reactive power at bus $i$. Traditionally, \cref{eq:pf-balance} is solved using iterative numerical methods, such as NR, which is susceptible
to the aforementioned limitations. 

$P_i$ and $Q_i$ can be expressed in rectangular coordinates:
\begin{subequations} \label{eq:pq_sum}
    \begin{align}
        P_i & = \sum_{k=1}^N G_{ik}(\mu_i\mu_k + \omega_i\omega_k) + B_{ik}(\omega_i\mu_k - \mu_i\omega_k), \label{eq:p-sum} \\
        Q_i & = \sum_{k=1}^N G_{ik}(\omega_i\mu_k - \mu_i\omega_k) - B_{ik}(\mu_i\mu_k + \omega_i\omega_k),  \label{eq:q-sum}
    \end{align}
\end{subequations}
with $\mu_i$ and $\omega_i$ being respectively the real and imaginary parts of the complex voltage at bus $i$; $G_{ik}$ the conductance and $B_{ik}$ the susceptance between bus $i$ and bus $k$.

To further prepare \cref{eq:pf-balance} for an Ising model, \cref{eq:pq_sum} can be re-ordered, as:
\begin{subequations} \label{eq:pq-expanded}
    \begin{align}
        P_i & = \sum_{k=1}^{N} \mu_i G_{ik} \mu_k + \omega_i G_{ik} \omega_k + \omega_i B_{ik} \mu_k - \mu_i B_{ik} \omega_k , \label{eq:p-expanded} \\
        Q_i & = \sum_{k=1}^{N} \omega_i G_{ik} \mu_k - \mu_i G_{ik} \omega_k - \mu_i B_{ik} \mu_k - \omega_i B_{ik} \omega_k .\label{eq:q-expanded}
    \end{align}
\end{subequations}

Next, the continuous variables 
\(\mu_i\) and \(\omega_i\) from \cref{eq:pq-expanded} must be discretized. One possible approach is to introduce multiple spin decision variables per bus $i$, each associated with a predefined increment. These spin variables determine whether the increment is added to or subtracted from a given base value for \(\mu_i\) and \(\omega_i\). While this method provides a fine-grained search space, it requires a large number of spin decision variables per bus $i$ for both \(\mu_i\) and \(\omega_i\), which significantly increases the dimensionality of the combinatorial optimization problem. 

An alternative, more efficient approach is to assign a single spin decision variable to each \(\mu_i\) and each \(\omega_i\), while iteratively refining their base values \(\mu_i^0\) and \(\omega_i^0\), that is:
\begin{subequations} \label{eq:discretization}
    \begin{align}
        \mu_i & := \mu_i^0 + s_{i}^\mu \, \Delta \mu_i, \label{eq:mu-discretization}\\
        \omega_i & := \omega_i^0 + s_{i}^\omega \, \Delta \omega_i, \label{eq:omega-discretization}
    \end{align}
\end{subequations}
where the spin decision variables, \(s_{i}^\mu \in \{\pm1\}^{N}, \ s_{i}^\omega \in \{\pm1\}^{N}\), determine whether the base values, \(\mu_i^0\) and \(\omega_i^0\), are increased or decreased per iteration. In doing so, \(\mu_i\) and \(\omega_i\) are then iteratively updated.

To further obtain a combinatorial optimization problem suitable for an Ising model, \cref{eq:pf-balance} is recast as the minimization of the sum of squared residuals for all terms:
\begin{equation} \label{eq:Hamiltonian}
    \min_{\mathbf{s} \in \{\pm 1\}^{2N}} \sum_{i=1}^{N} \big(P_i - P_i^{\mathrm{G}} + P_i^{\mathrm{D}}\big)^2 + \big(Q_i - Q_i^{\mathrm{G}} + Q_i^{\mathrm{D}}\big)^2 ,
\end{equation}
where $\mathbf{s}\in\{\pm1\}^{2N}$ is a vector of spin decision variables, that is, $s_{i}^{\{\mu,\omega\}}\ \forall i \in \{1,\dots,N\}$.

The increments, \(\Delta \mu_i\) and \(\Delta \omega_i\), are iteration-dependent and gradually decrease over time, thus allowing the optimization to transition from coarse exploration 
to fine refinement of the solution space:
{\small
\begin{subequations} \label{eq:increment}
    \begin{align}
        \Delta\mu_i&=\exp \left( \ln(0.1) + \frac{\text{it} \cdot \left( \ln(1\times10^{-4}) - \ln(0.1) \right)}{\text{it}_{\max}} \right), \label{eq:mu-increment}\\
        \Delta\omega_i&=\exp \left( \ln(0.05) + \frac{\text{it} \cdot \left( \ln(1\times10^{-5}) - \ln(0.05) \right)}{\text{it}_{\max}} \right), \label{eq:omega-increment}
    \end{align}
\end{subequations}
}
where `it' is the iteration counter, `$\text{it}_\text{max}$' is the maximum number of iterations. Since $\mu$ and $\omega$ are expressed in p.u., the maximum and minimum values for $\Delta\mu$ and $\Delta\omega$ are chosen accordingly to remain within numerically stable ranges.

Substituting \cref{eq:discretization} into \cref{eq:p-expanded} yields:
\begin{equation} \label{eq:p-ising}
\begin{split}
    {}& P_i= \sum_{k=1}^{N}
    \bigg[\mu_i^0 G_{ik} \mu_k^0
    + 
    \omega_i^0 G_{ik} \omega_k^0
    + 
    \omega_i^0 B_{ik} \mu_k^0
    - 
    \mu_i^0 B_{ik} \omega_k^0\bigg] \\
    &\quad
    +
    \bigg[\mu_i^0 G_{ik} s_{k}^\mu\Delta \mu_k
    +
    s_{i}^\mu\Delta \mu_i G_{ik} \mu_k^0
    +
    \omega_i^0 G_{ik} s_{k}^\omega\Delta \omega_k\\
    &\quad
    +
    s_{i}^\omega\Delta \omega_i G_{ik} \omega_k^0
    +
    \omega_i^0 B_{ik} s_{k}^\mu\Delta \mu_k
    + 
    s_{i}^\omega\Delta \omega_i B_{ik} \mu_k^0\\
    &\quad
    -
    \mu_i^0 B_{ik} s_{k}^\omega\Delta \omega_k
    - 
    s_{i}^\mu\Delta \mu_i B_{ik} \omega_k^0\bigg] \\
    &\quad
    +
    \bigg[s_{i}^\mu\Delta \mu_i G_{ik} s_{k}^\mu\Delta \mu_k
    + 
    s_{i}^\omega\Delta \omega_i G_{ik} s_{k}^\omega\Delta \omega_k\\
    &\quad
    + 
    s_{i}^\omega\Delta \omega_i B_{ik} s_{k}^\mu\Delta \mu_k
    - 
    s_{i}^\mu\Delta \mu_i B_{ik} s_{k}^\omega\Delta \omega_k\bigg],
\end{split}
\end{equation}
where the three bracketed expressions correspond, respectively, to constant, linear, and quadratic contributions to \(P_i\).

Similarly, substituting \cref{eq:discretization} into \cref{eq:q-expanded} yields:
\begin{equation} \label{eq:q-ising}
\begin{split}
    {}& Q_i= \sum_{k=1}^{N}
    \bigg[\omega_i^0 G_{ik} \mu_k^0
    - 
    \mu_i^0 G_{ik} \omega_k^0
    - 
    \mu_i^0 B_{ik} \mu_k^0
    - 
    \omega_i^0 B_{ik} \omega_k^0\bigg]\\
    &\quad
    +
    \bigg[\omega_i^0 G_{ik} s_{k}^\mu\Delta \mu_k
    + 
    s_{i}^\omega\Delta \omega_i G_{ik} \mu_k^0
    -
    \mu_i^0 G_{ik} s_{k}^\omega\Delta \omega_k\\
    &\quad
    - 
    s_{i}^\mu\Delta \mu_i G_{ik} \omega_k^0
    - 
    \mu_i^0 B_{ik} s_{k}^\mu\Delta \mu_k
    - 
    s_{i}^\mu\Delta \mu_i B_{ik} \mu_k^0\\
    &\quad
    -
    \omega_i^0 B_{ik} s_{k}^\omega\Delta \omega_k
    - 
    s_{i}^\omega\Delta \omega_i B_{ik} \omega_k^0\bigg]\\
    &\quad
    +
    \bigg[s_{i}^\omega\Delta \omega_i G_{ik} s_{k}^\mu\Delta \mu_k 
    - 
    s_{i}^\mu\Delta \mu_i G_{ik} s_{k}^\omega\Delta \omega_k\\
    &\quad
    - 
    s_{i}^\mu\Delta \mu_i B_{ik} s_{k}^\mu\Delta \mu_k 
    - 
    s_{i}^\omega\Delta \omega_i B_{ik} s_{k}^\omega\Delta \omega_k\bigg].
\end{split}
\end{equation}

In this study, we develop an Ising model for the proposed combinatorial PF analysis. An Ising model~\cite{Goto2019} is formulated as the problem of finding the spin decision variable vector $\mathbf{s} \in \{\pm 1\}^n$ that minimizes:
\begin{equation} \label{eq:ising}
    \min_{\mathbf{s} \in \{\pm 1\}^n} \sum_{i=1}^n h_i s_i + \sum_{\langle i,j\rangle} J_{ij} s_i s_j,
\end{equation}
where $h_i$ represents an external field that acts as a linear bias, influencing the tendency of $s_i$ toward $+1$ or $-1$, and $J_{ij}$ is the interaction coefficient between spins $i$ and $j$. $\langle i,j \rangle$ denotes all unique pairs of spins with $i<j$. 

Note that \cref{eq:p-ising,eq:q-ising} already contain quadratic terms resulting from the interactions between pairs of spin variables, e.g., $s_{i}^\mu\Delta \mu_i G_{ik} s_{k}^\mu\Delta \mu_k$ in \cref{eq:p-ising}. Therefore, substituting \cref{eq:p-ising,eq:q-ising} into \cref{eq:Hamiltonian} yields a fourth-order polynomial in the spin variables. To solve this minimization problem, we use two quantum computing paradigms: GQC and AQC. For GQC, we employ QAOA, which utilizes a PQC and variational optimization, thereby allowing direct encoding and optimization of problems with higher-order interactions. For AQC, we use two Ising machines: QA and OIIO. Since current Ising machines can only handle quadratic interactions, we use the Python package PyQUBO\footnote{\url{https://pyqubo.readthedocs.io}} to reduce higher-order interactions to quadratic ones. QIIO natively supports higher-order reduction, enabling the direct implementation of the fourth-order polynomial. 

The iterative scheme used for the combinatorial PF analysis is outlined in \cref{alg:aqpf}. First, the generation and demand power vectors, $\mathbf{P}^{\mathrm{G}}$, $\mathbf{P}^{\mathrm{D}}$, and $\mathbf{Q}^{\mathrm{D}}$, along with the admittance matrix $\mathbf{Y}$, are initialized according to the given power system data (lines 1-3). The increment vectors, $\Delta \mu$ and $\Delta \omega$, and the initial real and imaginary voltage vectors, $\mathbf{\mu}^0$ and $\mathbf{\omega}^0$, are then assigned by user-defined values (lines 4-5). Based on these initializations, the corresponding active and reactive power vectors, $\mathbf{P}$ and $\mathbf{Q}$, are computed (line 6), excluding the \emph{slack} bus entries, $P_1$ and $Q_1$, to remain consistent with the PF equations. Next, the problem Hamiltonian \cref{eq:Hamiltonian} is evaluated for the initial $\mathbf{\mu}^0$ and $\mathbf{\omega}^0$ (line 7), providing a first estimate of the solution. A convergence threshold $\epsilon$ is set and the iteration counter `it` is initialized to zero (line 8). During each iteration, the problem Hamiltonian is minimized with a given solver (line 11). In this study, three solvers are used: QAOA, QA, and QIIO. The resulting spin variable vector $\mathbf{s} \in \{\pm1\}^{2N}$ is used to update the voltage components $\mu_i$ and $\omega_i$ according to \cref{eq:discretization} (line 11). If \cref{eq:Hamiltonian} falls below $\epsilon$, the corresponding complex voltages, $\mathbf{\mu}+j\mathbf{\omega}$, are accepted as the PF solution. Otherwise, the base voltage values are reset, $\mathbf{\mu}^0:=\mu$ and $\mathbf{\omega}^0:=\omega$ (line 14), and the increments are adjusted using \cref{eq:increment}. The loop continues with the updated Hamiltonian until the algorithm converges to a solution. 

\begin{algorithm}[t]
\caption{Iterative Scheme for Combinatorial Power Flow Analysis.}\label{alg:aqpf}
\begin{algorithmic}[1]
    \State Initialize generation vector $\mathbf{P}^{\mathrm{G}} = [P^{\mathrm{G}}_1, P^{\mathrm{G}}_2, \dots, P^{\mathrm{G}}_{N_\mathrm{G}}]$
    \State Initialize demand vectors $\mathbf{P}^{\mathrm{D}} = [P^{\mathrm{D}}_1, \dots, P^{\mathrm{D}}_{N - N_\mathrm{G} - 1}]$ and $\mathbf{Q}^{\mathrm{D}} = [Q^{\mathrm{D}}_1, \dots, Q^{\mathrm{D}}_{N - N_\mathrm{G} - 1}]$
    \State Initialize the admittance matrix $\mathbf{Y}= \{G_{ik} + jB_{ik}: i,k=1,\dots,N\}$
    \State Initialize increment vectors $\Delta \mu \gets 0.1$, $\Delta \omega \gets 0.05$
    \State Initialize voltage vectors $\mathbf{\mu}^0 = [1, \dots, 1]$, $\mathbf{\omega}^0 = [0, \dots, 0]$
    \State Compute initial active and reactive power vectors $\mathbf{P} = [P_2, \dots, P_N]$ and $\mathbf{Q} = [Q_2, \dots, Q_N]$ using \cref{eq:pq-expanded}
    \State Evaluate \cref{eq:Hamiltonian} with $\mathbf{\mu}^0$ and $\mathbf{\omega}^0$
    \State Set convergence threshold $\epsilon \gets 1 \times 10^{-3}$ and iteration counter $\text{it} \gets 0$
    \While{\cref{eq:Hamiltonian}$ > \epsilon$ and $\text{it} < \text{it}_\text{max}$}
        \State Minimize \cref{eq:Hamiltonian} with a given solver
        \State Update voltage vectors $\mathbf{\mu}$ and $\mathbf{\omega}$ using \cref{eq:discretization}
        \State Calculate $\mathbf{P}$ and $\mathbf{Q}$ with updated $\mathbf{\mu}$ and $\mathbf{\omega}$ using \cref{eq:pq-expanded}
        \State Evaluate \cref{eq:Hamiltonian} updated $\mathbf{\mu}$ and $\mathbf{\omega}$
        \State Reset base voltage vectors $\mathbf{\mu}^0 := \mu$ and $\mathbf{\omega}^0 := \omega$
        \State Update increment vectors $\Delta \mu$ and $\Delta \omega$ using \cref{eq:increment}
        \State Increment iteration counter: $\text{it} \gets \text{it} + 1$
    \EndWhile
    \State Return complex voltage solution: $\mathbf{V} = \mathbf{\mu} + j \mathbf{\omega}$
\end{algorithmic}
\end{algorithm}

\section{Quantum Approximate Optimization Algorithm}
QAOA belongs to the GQC paradigm and is designed for solving combinatorial optimization problems. QAOA encodes problems as an Ising model or an equivalent QUBO, and the goal is to minimize a cost Hamiltonian defined over spin/binary variables. QAOA uses a parameterized sequence of discrete quantum gates that alternate between applying a cost Hamiltonian and a mixer Hamiltonian. Note that the cost Hamiltonian encodes the objective function of the given combinatorial optimization problem, while the mixer Hamiltonian drives transitions between computational basis states to ensure exploration of the solution space rather than getting stuck in a single configuration. By optimizing the gate parameters classically, the algorithm prepares a quantum state with a high probability of yielding near-optimal solutions when measured. The algorithm's depth controls the trade-off between solution quality and circuit complexity. While a higher depth can improve approximations, it also increases circuit complexity and may hinder the convergence of the classical optimizer.

QAOA has been experimentally demonstrated on near-term GQC hardware, such as superconducting qubit platforms (e.g., \cite{Weidenfeller2022}). While current devices are constrained by the number of qubits and gate fidelity, they are flexible in encoding an arbitrary cost Hamiltonian without the embedding restrictions of AQC. In principle, higher-order interactions can be incorporated into the cost Hamiltonian, though often at the expense of additional qubits or circuit depth. The performance of QAOA depends critically on both the quality of the quantum hardware and the classical optimizer that tunes the variational parameters to maximize the solution probability.

\section{Quantum Annealing}
Quantum annealing is a metaheuristic for solving combinatorial optimization problems by exploiting quantum-mechanical effects. Similar to classical simulated annealing, it seeks to find the global minimum of an energy landscape; however, it uses quantum tunneling to escape local minima more efficiently. A combinatorial optimization problem is typically formulated as an Ising model or an equivalent QUBO representation. The system is composed of interacting qubits that encode the problem variables. It evolves according to a time-dependent problem Hamiltonian, starting from an initial superposition of all possible states and gradually converging to the ground state, which encodes the optimal solution. This evolution is governed by the adiabatic theorem, which ensures that sufficiently slow changes in the problem Hamiltonian keep the system in its ground state~\cite{Goto2019,Lucas2014}.

Specialized hardware implementations of quantum annealing, also known as Ising machines, have been developed. QA is among the most prominent examples, featuring over 5,000 qubits and 35,000 couplers, in which qubits interact according to programmable coefficients that define the problem Hamiltonian, with connectivity constraints requiring careful mapping of the logical problem graph onto the hardware topology. These limitations make embedding a critical step, as it can introduce overhead and reduce effective precision. DA is another example. It represents an application-specific complementary metal-oxide semiconductor (CMOS) hardware\footnote{\url{www.fujitsu.com/global/services/business-services/digital-annealer}} that emulates annealing behavior at room temperature. Unlike QA, which relies on cryogenic qubits, DA supports massively parallel simulated annealing and allows full connectivity between binary decision variables. Its latest generation, i.e., QIIO, can handle tens of thousands of binary decision variables with high numerical precision and introduces advanced techniques, such as parallel tempering, which runs multiple replicas at different temperatures to avoid local minima~\cite{Yin2023}.

In this study, the implementation is based on spin decision variables. Therefore, the binary bitstrings obtained from QIIO are converted into spin representations using the standard transformation $s_i = 2x_i -1$, where $s_i$ and $x_i$ denote the spin and binary variables, respectively.

\section{Results}
We propose the combinatorial PF analysis, based on which an Ising model is developed. The problem Hamiltonian is solved using the two quantum computing paradigms: GQC and AQC. Experiments are conducted on a standard 4-bus test system consisting of one \emph{slack} bus and three \emph{load} buses. For a given load scenario, the combinatorial PF analysis is run across three solvers: QA, QIIO, and QAOA.

\subsection{Model Setup}
For the GQC experiments, we implement QAOA using PennyLane's \texttt{lightning.qubit} statevector simulator, a high-performance statevector backend that efficiently simulates mid-scale quantum circuits. The QAOA ansatz with $p$ alternating layers of cost and mixer Hamiltonians/unitaries is defined as:
\begin{equation}
\ket{\psi(\boldsymbol{\gamma}, \boldsymbol{\beta})} = \prod_{k=1}^{p} e^{-i \beta_k H_M} e^{-i \gamma_k H_C} \ket{+}^{\otimes 2N},
\end{equation}
where $H_M = \sum_{i=1}^{2N} X_i$ is the mixer Hamiltonian with Pauli-$X$ operators acting on each qubit, and $H_C$ is the problem-specific cost Hamiltonian defined in \cref{eq:Hamiltonian}. The system size is $2N$, where $N$ is the number of buses in the PF equations \cref{eq:pq-expanded}, since both $\mu$ and $\omega$ variables are discretized. The parameters $\boldsymbol{\gamma} = (\gamma_1, \dots, \gamma_p)$ and $\boldsymbol{\beta} = (\beta_1, \dots, \beta_p)$ are randomly initialized within $[0,2\pi]$, and optimized iteratively to minimize the expected energy $\langle H_C \rangle$. Each expectation value is estimated from projective measurements in the computational basis, with 1,000 shots per evaluation.

The optimization loop is classical. \texttt{Adam} optimizer is used to update parameters over 100 steps, with a learning rate of $0.1$. For each step, the ansatz $\ket{\psi(\boldsymbol{\gamma}, \boldsymbol{\beta})}$ is prepared, measured, and $\langle H_C \rangle$ is computed. The optimizer then updates the parameters in order to converge towards an approximate ground state of $H_C$. This implementation corresponds directly to line 10 in \cref{alg:aqpf}, where the variational quantum subroutine is called within the iterative scheme for the combinatorial PF analysis. Table \ref{tab:qaoa_detail} summarizes the QAOA hyperparameters used for the 4-bus test system experiment. The specifications are selected based on preliminary runs to balance convergence accuracy and computational cost.

\begin{table}[h]
\centering
\caption{QAOA implementation parameters for the combinatorial PF analysis based on the 4-bus test system.}
\label{tab:qaoa_detail}
\footnotesize
\renewcommand{\arraystretch}{1.3}
    \begin{tabular}{lc}
        \toprule
        Parameter & Specification \\
        \midrule
        Number of qubits $q$                        & $8 \ (-)$\\
        Circuit depth $p$                       & $2 \ (-)$\\
        Optimization steps $st$                 & $100 \ (-)$\\
        Learning rate $lr$                      & $0.1 \ (-)$\\
        Shots per expectation evaluation $e$ & $1000 \ (-)$\\
        Convergence threshold  $\epsilon$ & $1 \times 10^{-3} \ (-)$\\
        Optimizer & \texttt{Adam}\\
        Simulator backend & \texttt{lightning.qubit}\\
        \bottomrule
    \end{tabular}
\end{table}

For the AQC experiments, two Ising machines are used, i.e., QA and QIIO. This implementation directly corresponds to line 10 in \cref{alg:aqpf}, where the quantum/digital annealer is called. For QA, the problem Hamiltonian is mapped to the QA hardware graph using minor embedding. The chain strength is tuned to balance between preventing chain breaks and preserving the weight hierarchy of the quadratic interactions. For both approaches, 1,000 readouts/samples are collected from the annealing run, and the sample with the lowest energy is selected as the minimized solution to \cref{eq:Hamiltonian}. Table \ref{tab:qa_detail} summarizes the parameters used for QA and QIIO based on the 4-bus test system. Detailed information about the quantum annealing implementation can be found in our prior studies, e.g.,~\cite{kaseb2024power,kaseb2025aqpf2,Kaseb2025SolvingComputing}.

\begin{table}[h]
\centering
\caption{Quantum annealing implementation parameters for the combinatorial PF analysis based on the 4-bus test system.}
\label{tab:qa_detail}
\renewcommand{\arraystretch}{1.3}
    \begin{tabular}{lc}
        \toprule
        Parameter & Specification \\
        \midrule
        QA Number of qubits   $q$     & $26 \ (-)$\\
        QIIO Number of qubits $q$    & $20 \ (-)$\\
        Number of readouts $r$      & $1000 \ (-)$\\
        Convergence threshold $\epsilon$ & $1 \times 10^{-3} \ (-)$\\
        QIIO Time limit & $10 \ (\text{seconds})$\\
        QIIO Precision & $64-\text{bits}$\\
        QIIO Overall timeout & $3600 \ (\text{seconds})$\\
        QA Chip ID  & \texttt{Advantage2\_system1.5}\\
        QA Minor embedding & \texttt{EmbeddingComposite}\\
        \bottomrule
    \end{tabular}
\end{table}

\subsection{Model Performance}
\cref{tab:computational_detail} summarizes the computational details of solving the combinatorial PF analysis using QA, QIIO, and QAOA based on the 4-bus test system. The active and reactive power demands for \emph{load} buses, $\mathbf{P}^{\mathrm{D}}$ and $\mathbf{Q}^{\mathrm{D}}$, and $\mu_0$ and $\omega_0$ for the \emph{slack} bus are specified, while $\mu_i$ and $\omega_i$ are unknown for all \emph{load} buses $\forall i \in \{1,2,3\}$. 

The Ising model implementation results in 26 spin variables for QA, where higher-order interactions are reduced to quadratic terms using PyQUBO, and 20 decision variables for QIIO. The discrepancy in the number of variables reflects the different strategies of the underlying software frameworks in handling higher-order interactions. For QAOA, the variable count corresponds to the number of qubits. With two variables per bus $i$ (one for $\mu_i$ and one for $\omega_i$), the 4-bus test system requires 8 qubits. Compilation times for QA and QIIO are comparable, as both involve classical preprocessing and reduction of higher-order terms. QAOA, in contrast, exhibits a compile time that is one order of magnitude higher, primarily due to circuit transpilation and parameter initialization, as shown in \cref{tab:computational_detail}.

Convergence behavior differs considerably across solvers. QA requires 222 iterations to satisfy the tolerance $\epsilon = 1 \times 10^{-3}$, with an average iteration time of 0.015 seconds (QPU access time). Note that the wall-clock time per iteration is, however, 1.25 seconds, reflecting overheads from minor embedding and QPU communication (programming, annealing, readout, and sampling). QIIO achieves convergence in 63 iterations with the average iteration time of 0.06 seconds, excluding communication overhead. QAOA, executed on a simulator, is dominated by repeated circuit evaluations and optimizer updates (100 steps per run). Despite 300 iterations, it does not reach the predefined threshold.

\begin{table}[h]
\centering
\caption{Computational detail for the combinatorial PF analysis solved with QA, QIIO, and QAOA based on the 4-bus test system.}
\label{tab:computational_detail}
\renewcommand{\arraystretch}{1.3}
    \begin{tabular}{lccccc}
        \toprule
        Solver & \# of & Compile & \# of & Time per & Residual \\
        & Variables & Time [s] & Iterations & Iteration [s] & [-] \\
        \midrule
        QA   & 26 & 0.003 & 222 & 0.015 & $5.18 \times 10^{-4}$ \\
        QIIO & 20 & 0.025  & 63  & 0.06  & $3.31 \times 10^{-4}$ \\
        QAOA & 8  & 0.03  & 300 & 15.6  & $2.49 \times 10^{-3}$ \\
        \bottomrule
    \end{tabular}
\end{table}

\cref{tab:pf_results} summarizes the bus complex voltages ($\mu_i + j\omega_i$) obtained from the combinatorial PF analysis using QA, QIIO, and QAOA, in comparison with the NR method for the 4-bus test system. The \emph{slack} bus ($i=0$) with fixed values $\mu_0=1$ and $\omega_0=0$ is not shown. Both QA and QIIO reproduce the NR solution with high accuracy, with deviations on the order of $10^{-3}$ for both $\vec{\mu}$ and $\vec{\omega}$. QAOA achieves comparable accuracy for $\vec{\mu}$, but shows slightly larger deviations for $\vec{\omega}$ and does not converge to the predefined threshold within 300 iterations.

\begin{table}[h] 
\caption{Performance comparison of QA, QIIO, and QAOA with the NR solver for the 4-bus test system. The \emph{slack} bus $i=0$ with known $\mu_0=1$ and $\omega_0=0$ is not shown.} 
\label{tab:pf_results}
\centering
\renewcommand{\arraystretch}{1.3}
    \begin{tabular}{lcccccc}
        \toprule
         & $\mu_{1}$ & $\mu_{2}$ & $\mu_{3}$ & $\omega_{1}$ & $\omega_{2}$ & $\omega_{3}$ \\
        \midrule
        NR    & 0.902 & 0.916 & 0.890 & -0.092 & -0.080 & -0.104 \\
        QA    & 0.901 & 0.915 & 0.889 & -0.093 & -0.080 & -0.105 \\
        QIIO  & 0.901 & 0.915 & 0.889 & -0.092 & -0.080 & -0.105 \\
        QAOA  & 0.902 & 0.916 & 0.890 & -0.089 & -0.078 & -0.099 \\
        \bottomrule
    \end{tabular}
\end{table}

\begin{figure}[t]
\centering
\includegraphics[width=6.4in]{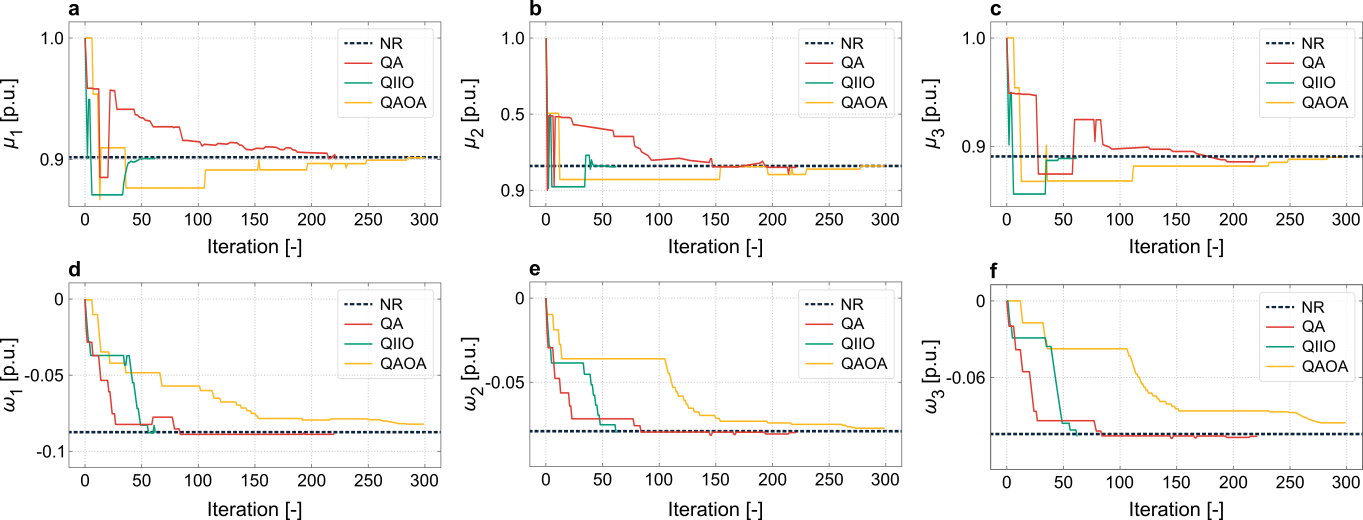}
\caption{Representation of $\mathbf{\mu} = [\mu_1, \mu_2, \mu_3]$ and $\mathbf{\omega} = [\omega_1, \omega_2, \omega_3]$ obtained by QA, QIIO, and QAOA for the 4-bus test system. The \emph{slack} bus $i=0$ is not shown. The graphs include $\mu_i$ and $\omega_i$ obtained from the NR solver.}
\label{fig:pf_results}
\end{figure}

\cref{fig:pf_results} shows the evolution of bus complex voltage $\mathbf{\mu} = [\mu_1, \mu_2, \mu_3]$ and $\mathbf{\omega} = [\omega_1, \omega_2, \omega_3]$ obtained with QA, QIIO, and QAOA for the 4-bus test system over iterations. The \emph{slack} bus values $\mu_0=1$ and $\omega_0=0$ are not shown. Results from the NR are included as references. \cref{fig:pf_results} (a–c) show the convergence of $\mu_{1-3}$, where $\mu_2$ reaches its NR value faster than $\mu_1$ and $\mu_3$. \cref{fig:pf_results} (d–f) display the corresponding $\omega_{1-3}$, with $\omega_2$ converging more rapidly than $\omega_1$ and $\omega_3$. QA oscillates around the NR values for $\omega_{1-3}$ before stabilizing. Its longer overall convergence rate is mainly due to the slower convergence of $\mu_1$ and $\mu_3$. Among the methods, QIIO converges the fastest, while QAOA is the slowest. Nevertheless, all three approaches ultimately yield solutions consistent with the NR benchmark.

\section{Discussion}
Quantum computing is an emerging field in which both hardware and software are undergoing rapid development. Although recent advances in quantum hardware have progressed faster than anticipated only a few years ago, current devices are not yet capable of supporting large-scale and practical applications. Nevertheless, without well-suited algorithms and software frameworks, even fully developed quantum processors would have limited practical value. Consequently, research on quantum algorithms for application domains where computational performance is critical must progress in parallel with hardware development. At present, despite the availability of early quantum hardware, electricity grid operations cannot meaningfully benefit from it because quantum algorithms tailored to power system problems are still in development. Motivated by this gap, this paper investigates two major quantum computing paradigms, GQC and AQC, for solving the PF problem. Based on the results presented in this work, the following observations can be made:
\begin{itemize}
    \item On real quantum hardware, i.e., QA, repeated executions may yield different results because device conditions, such as noise levels, calibration, or qubit coherence, can vary over time. In addition, QA is prone to disconnection. For example, we repeatedly encountered the error \textit{“Remote end closed connection without response”} when using \texttt{Advantage2\_system1.5} and \texttt{Advantage\_system6.4} for larger test systems beyond the 4-bus test system.
    \item For the standard 4-bus test system, we conducted experiments on both \texttt{Advantage2\_system1.5} and \texttt{Advantage} \texttt{\_system6.4}. The former proved to be up to $20\%$ faster per iteration and consistently produced better results than the latter.
    \item The choice of a small test system reflects current limitations in GQC and the associated computational cost. In contrast, AQC approaches can already address larger problem sizes. For example, power systems with up to 1354 buses have been solved using DA in our previous work~\cite{kaseb2025aqpf2}.  
\end{itemize}

\section{Conclusion}
In this study, we implement combinatorial PF analysis on both GQC and AQC hardware. The PF problem is reformulated as an Ising model, which is then solved using different quantum solvers. For GQC, QAOA is implemented using PennyLane’s \texttt{lightning.qubit} statevector simulator. For AQC, both QA and QIIO are employed. Experiments are conducted on a standard 4-bus test system consisting of one \emph{slack} bus and three \emph{load} buses. The results are evaluated in terms of solution accuracy and computational performance, and indicate that both QA and QIIO converge within the predefined threshold before reaching the maximum iteration limit, with QIIO achieving the fastest performance in terms of iteration count. In contrast, QAOA exhibits limited accuracy under the tested configuration. However, QA, QIIO, and QAOA all produce solutions that closely approximate the NR benchmark, confirming that the proposed combinatorial PF formulation is consistent with the classical AC PF equations.

\section*{Acknowledgment}
This work is part of the DATALESs project (project no. 482.20.602), jointly financed by the Netherlands Organization for Scientific Research (NWO) and the National Natural Science Foundation of China (NSFC).
The authors gratefully acknowledge Heinz Wilkening and the service contract "Quantum Computing for Load Flow" (contract no. 690523) with the European Commission Directorate-General Joint Research Centre (EC DG JRC). 
The authors also like to thank Fujitsu Technology Solutions for providing access to the QIIO software\footnote{\url{https://en-portal.research.global.fujitsu.com/kozuchi}}.
Furthermore, the authors acknowledge TNO for the access to TNO's Quantum Application Lab Facility, with special thanks to Frank Phillipson for his assistance.

\printcredits

\bibliographystyle{cas-model2-names}
\bibliography{cas-refs}

\begin{thebibliography}{27}
\expandafter\ifx\csname natexlab\endcsname\relax\def\natexlab#1{#1}\fi
\providecommand{\url}[1]{\texttt{#1}}
\providecommand{\href}[2]{#2}
\providecommand{\path}[1]{#1}
\providecommand{\DOIprefix}{doi:}
\providecommand{\ArXivprefix}{arXiv:}
\providecommand{\URLprefix}{URL: }
\providecommand{\Pubmedprefix}{pmid:}
\providecommand{\doi}[1]{\href{http://dx.doi.org/#1}{\path{#1}}}
\providecommand{\Pubmed}[1]{\href{pmid:#1}{\path{#1}}}
\providecommand{\bibinfo}[2]{#2}
\ifx\xfnm\relax \def\xfnm[#1]{\unskip,\space#1}\fi
\bibitem[{Ali et~al.(2025)Ali, Wadho, Talpur, Talpur, Alshudukhi, Humayun, Talpur, Mamun, Naseem, Abro, Talpur and Shah}]{Ali2025}
\bibinfo{author}{Ali, S.}, \bibinfo{author}{Wadho, S.A.}, \bibinfo{author}{Talpur, K.R.}, \bibinfo{author}{Talpur, B.A.}, \bibinfo{author}{Alshudukhi, K.S.}, \bibinfo{author}{Humayun, M.}, \bibinfo{author}{Talpur, S.R.}, \bibinfo{author}{Mamun, M.A.A.}, \bibinfo{author}{Naseem, M.}, \bibinfo{author}{Abro, A.}, \bibinfo{author}{Talpur, D.B.}, \bibinfo{author}{Shah, A.}, \bibinfo{year}{2025}.
\newblock \bibinfo{title}{Next-generation quantum security: The impact of quantum computing on cybersecurity—threats, mitigations, and solutions}.
\newblock \bibinfo{journal}{Computers \& Electrical Engineering} \bibinfo{volume}{128}, \bibinfo{pages}{110649}.
\newblock \DOIprefix\doi{10.1016/j.compeleceng.2025.110649}.
\bibitem[{Armas et~al.(2024)Armas, Creemers and Deleplanque}]{Armas2024}
\bibinfo{author}{Armas, L.F.P.}, \bibinfo{author}{Creemers, S.}, \bibinfo{author}{Deleplanque, S.}, \bibinfo{year}{2024}.
\newblock \bibinfo{title}{Solving the resource constrained project scheduling problem with quantum annealing}.
\newblock \bibinfo{journal}{Scientific Reports} \bibinfo{volume}{14}, \bibinfo{pages}{16784}.
\newblock \DOIprefix\doi{10.1038/s41598-024-67168-6}.
\bibitem[{Arrillaga and Smith(1998)}]{Arrillaga1998}
\bibinfo{author}{Arrillaga, J.}, \bibinfo{author}{Smith, B.}, \bibinfo{year}{1998}.
\newblock \bibinfo{title}{{ AC-DC power system analysis}}.
\newblock \bibinfo{publisher}{Institution of Electrical Engineers}, \bibinfo{address}{London}.
\bibitem[{Au-Yeung et~al.(2023)Au-Yeung, Chancellor and Halffmann}]{Au-Yeung2023}
\bibinfo{author}{Au-Yeung, R.}, \bibinfo{author}{Chancellor, N.}, \bibinfo{author}{Halffmann, P.}, \bibinfo{year}{2023}.
\newblock \bibinfo{title}{{NP}-hard but no longer hard to solve? using quantum computing to tackle optimization problems}.
\newblock \bibinfo{journal}{Frontiers in Quantum Science and Technology} \bibinfo{volume}{2}.
\newblock \DOIprefix\doi{10.3389/frqst.2023.1128576}.
\bibitem[{Basso et~al.(2022)Basso, Farhi, Marwaha, Villalonga and Zhou}]{Farhi2022}
\bibinfo{author}{Basso, J.}, \bibinfo{author}{Farhi, E.}, \bibinfo{author}{Marwaha, K.}, \bibinfo{author}{Villalonga, B.}, \bibinfo{author}{Zhou, L.}, \bibinfo{year}{2022}.
\newblock \bibinfo{title}{The quantum approximate optimization algorithm at high depth for maxcut on large-girth regular graphs and the sherrington-kirkpatrick model}, in: \bibinfo{booktitle}{17th Conference on the Theory of Quantum Computation, Communication and Cryptography}, \bibinfo{address}{Dagstuhl, Germany}. p.~\bibinfo{pages}{7}.
\bibitem[{Bienstock and Verma(2019)}]{Bienstock2019}
\bibinfo{author}{Bienstock, D.}, \bibinfo{author}{Verma, A.}, \bibinfo{year}{2019}.
\newblock \bibinfo{title}{{Strong {NP}-hardness of AC power flows feasibility}}.
\newblock \bibinfo{journal}{Operations Research Letters} \bibinfo{volume}{47}, \bibinfo{pages}{494--501}.
\newblock \DOIprefix\doi{10.1016/j.orl.2019.08.009}.
\bibitem[{Dupont et~al.(2023)Dupont, Evert, Hodson, Sundar, Jeffrey, Yamaguchi, Feng, Maciejewski, Hadfield, Alam, Wang, Grabbe, Lott, Rieffel, Venturelli and Reagor}]{Dupont2023}
\bibinfo{author}{Dupont, M.}, \bibinfo{author}{Evert, B.}, \bibinfo{author}{Hodson, M.J.}, \bibinfo{author}{Sundar, B.}, \bibinfo{author}{Jeffrey, S.}, \bibinfo{author}{Yamaguchi, Y.}, \bibinfo{author}{Feng, D.}, \bibinfo{author}{Maciejewski, F.B.}, \bibinfo{author}{Hadfield, S.}, \bibinfo{author}{Alam, M.S.}, \bibinfo{author}{Wang, Z.}, \bibinfo{author}{Grabbe, S.}, \bibinfo{author}{Lott, P.A.}, \bibinfo{author}{Rieffel, E.G.}, \bibinfo{author}{Venturelli, D.}, \bibinfo{author}{Reagor, M.J.}, \bibinfo{year}{2023}.
\newblock \bibinfo{title}{Quantum-enhanced greedy combinatorial optimization solver}.
\newblock \bibinfo{journal}{Science Advances} \bibinfo{volume}{9}.
\newblock \DOIprefix\doi{10.1126/sciadv.adi0487}.
\bibitem[{Ganesan et~al.(2021)Ganesan, Sobhana, Anuradha, Yellamma, Devi, Prakash and Naren}]{Ganesan2021}
\bibinfo{author}{Ganesan, V.}, \bibinfo{author}{Sobhana, M.}, \bibinfo{author}{Anuradha, G.}, \bibinfo{author}{Yellamma, P.}, \bibinfo{author}{Devi, O.R.}, \bibinfo{author}{Prakash, K.B.}, \bibinfo{author}{Naren, J.}, \bibinfo{year}{2021}.
\newblock \bibinfo{title}{Quantum inspired meta-heuristic approach for optimization of genetic algorithm}.
\newblock \bibinfo{journal}{Computers \& Electrical Engineering} \bibinfo{volume}{94}, \bibinfo{pages}{107356}.
\newblock \DOIprefix\doi{10.1016/j.compeleceng.2021.107356}.
\bibitem[{Goto et~al.(2019)Goto, Tatsumura and Dixon}]{Goto2019}
\bibinfo{author}{Goto, H.}, \bibinfo{author}{Tatsumura, K.}, \bibinfo{author}{Dixon, A.R.}, \bibinfo{year}{2019}.
\newblock \bibinfo{title}{Combinatorial optimization by simulating adiabatic bifurcations in nonlinear hamiltonian systems}.
\newblock \bibinfo{journal}{Science Advances} \bibinfo{volume}{5}.
\newblock \DOIprefix\doi{10.1126/sciadv.aav2372}.
\bibitem[{Grainger and Stevenson~Jr.(1994)}]{Grainger1994}
\bibinfo{author}{Grainger, J.J.}, \bibinfo{author}{Stevenson~Jr., W.D.}, \bibinfo{year}{1994}.
\newblock \bibinfo{title}{{Power System Analysis}}.
\newblock \bibinfo{publisher}{McGraw-Hill, Inc.}
\bibitem[{{Hadi Saadat}(1999)}]{Saadat1999}
\bibinfo{author}{{Hadi Saadat}}, \bibinfo{year}{1999}.
\newblock \bibinfo{title}{{Power system analysis}}.
\newblock \bibinfo{publisher}{McGraw Hill}, \bibinfo{address}{Singapore}.
\bibitem[{Imre(2014)}]{Imre2014}
\bibinfo{author}{Imre, S.}, \bibinfo{year}{2014}.
\newblock \bibinfo{title}{Quantum computing and communications – introduction and challenges}.
\newblock \bibinfo{journal}{Computers \& Electrical Engineering} \bibinfo{volume}{40}, \bibinfo{pages}{134--141}.
\newblock \DOIprefix\doi{10.1016/j.compeleceng.2013.10.008}.
\bibitem[{Jing et~al.(2023)Jing, Wang and Li}]{Jing2023}
\bibinfo{author}{Jing, H.}, \bibinfo{author}{Wang, Y.}, \bibinfo{author}{Li, Y.}, \bibinfo{year}{2023}.
\newblock \bibinfo{title}{Data-driven quantum approximate optimization algorithm for power systems}.
\newblock \bibinfo{journal}{Communications Engineering} \bibinfo{volume}{2}, \bibinfo{pages}{12}.
\newblock \DOIprefix\doi{10.1038/s44172-023-00061-8}.
\bibitem[{Kaseb et~al.(2024a)Kaseb, M{\"o}ller, Balducci, Palensky and Vergara}]{kaseb2024quantum}
\bibinfo{author}{Kaseb, Z.}, \bibinfo{author}{M{\"o}ller, M.}, \bibinfo{author}{Balducci, G.T.}, \bibinfo{author}{Palensky, P.}, \bibinfo{author}{Vergara, P.P.}, \bibinfo{year}{2024}a.
\newblock \bibinfo{title}{Quantum neural networks for power flow analysis}.
\newblock \bibinfo{journal}{Electric Power Systems Research} \bibinfo{volume}{235}, \bibinfo{pages}{110677}.
\newblock \DOIprefix\doi{10.1016/j.epsr.2024.110677}.
\bibitem[{Kaseb et~al.(2025a)Kaseb, M{\"{o}}ller, Palensky and Vergara}]{kaseb2025aqpf2}
\bibinfo{author}{Kaseb, Z.}, \bibinfo{author}{M{\"{o}}ller, M.}, \bibinfo{author}{Palensky, P.}, \bibinfo{author}{Vergara, P.P.}, \bibinfo{year}{2025}a.
\newblock \bibinfo{title}{Quantum-enhanced power flow and optimal power flow based on combinatorial reformulation}.
\newblock \bibinfo{journal}{arXiv preprint arXiv:2505.15978} .
\bibitem[{Kaseb et~al.(2025b)Kaseb, Moller, Palensky and Vergara}]{Kaseb2025SolvingComputing}
\bibinfo{author}{Kaseb, Z.}, \bibinfo{author}{Moller, M.}, \bibinfo{author}{Palensky, P.}, \bibinfo{author}{Vergara, P.P.}, \bibinfo{year}{2025}b.
\newblock \bibinfo{title}{{Solving Power System Problems using Adiabatic Quantum Computing}}.
\newblock \bibinfo{journal}{arXiv preprint arXiv:2504.06458} .
\bibitem[{Kaseb et~al.(2024b)Kaseb, M{\"o}ller, Vergara and Palensky}]{kaseb2024power}
\bibinfo{author}{Kaseb, Z.}, \bibinfo{author}{M{\"o}ller, M.}, \bibinfo{author}{Vergara, P.P.}, \bibinfo{author}{Palensky, P.}, \bibinfo{year}{2024}b.
\newblock \bibinfo{title}{Power flow analysis using quantum and digital annealers: a discrete combinatorial optimization approach}.
\newblock \bibinfo{journal}{Scientific Reports} \bibinfo{volume}{14}, \bibinfo{pages}{23216}.
\newblock \DOIprefix\doi{10.1038/s41598-024-73512-7}.
\bibitem[{Lucas(2014)}]{Lucas2014}
\bibinfo{author}{Lucas, A.}, \bibinfo{year}{2014}.
\newblock \bibinfo{title}{Ising formulations of many {NP} problems}.
\newblock \bibinfo{journal}{Frontiers in Physics} \bibinfo{volume}{2}.
\newblock \DOIprefix\doi{10.3389/fphy.2014.00005}.
\bibitem[{Mokryani et~al.(2016)Mokryani, Majumdar and Pal}]{Mokryani2016}
\bibinfo{author}{Mokryani, G.}, \bibinfo{author}{Majumdar, A.}, \bibinfo{author}{Pal, B.C.}, \bibinfo{year}{2016}.
\newblock \bibinfo{title}{{Probabilistic method for the operation of three‐phase unbalanced active distribution networks}}.
\newblock \bibinfo{journal}{IET Renewable Power Generation} \bibinfo{volume}{10}, \bibinfo{pages}{944--954}.
\newblock \DOIprefix\doi{10.1049/iet-rpg.2015.0334}.
\bibitem[{Pareek et~al.()Pareek, Jayakumar, Coffrin and Misra}]{PareekDemystifyingAdvantage}
\bibinfo{author}{Pareek, P.}, \bibinfo{author}{Jayakumar, A.}, \bibinfo{author}{Coffrin, C.}, \bibinfo{author}{Misra, S.}, .
\newblock \bibinfo{title}{{Demystifying Quantum Power Flow: Unveiling the Limits of Practical Quantum Advantage}}.
\newblock \bibinfo{type}{Technical Report}.
\bibitem[{Pelofske et~al.(2023)Pelofske, B{\"a}rtschi and Eidenbenz}]{Pelofske2023}
\bibinfo{author}{Pelofske, E.}, \bibinfo{author}{B{\"a}rtschi, A.}, \bibinfo{author}{Eidenbenz, S.}, \bibinfo{year}{2023}.
\newblock \bibinfo{title}{Quantum annealing vs. qaoa: 127 qubit higher-order ising problems on nisq computers}, in: \bibinfo{editor}{Bhatele, A.}, \bibinfo{editor}{Hammond, J.}, \bibinfo{editor}{Baboulin, M.}, \bibinfo{editor}{Kruse, C.} (Eds.), \bibinfo{booktitle}{High Performance Computing}, \bibinfo{publisher}{Springer Nature Switzerland}. pp. \bibinfo{pages}{240--258}.
\bibitem[{Pelofske et~al.(2024)Pelofske, Bärtschi and Eidenbenz}]{Pelofske2024}
\bibinfo{author}{Pelofske, E.}, \bibinfo{author}{Bärtschi, A.}, \bibinfo{author}{Eidenbenz, S.}, \bibinfo{year}{2024}.
\newblock \bibinfo{title}{Short-depth qaoa circuits and quantum annealing on higher-order ising models}.
\newblock \bibinfo{journal}{npj Quantum Information} \bibinfo{volume}{10}, \bibinfo{pages}{30}.
\newblock \DOIprefix\doi{10.1038/s41534-024-00825-w}.
\bibitem[{Tostado-Véliz et~al.(2021)Tostado-Véliz, Hasanien, Turky, Alkuhayli, Kamel and Jurado}]{Tostado2021}
\bibinfo{author}{Tostado-Véliz, M.}, \bibinfo{author}{Hasanien, H.M.}, \bibinfo{author}{Turky, R.A.}, \bibinfo{author}{Alkuhayli, A.}, \bibinfo{author}{Kamel, S.}, \bibinfo{author}{Jurado, F.}, \bibinfo{year}{2021}.
\newblock \bibinfo{title}{Mann-iteration process for power flow calculation of large-scale ill-conditioned systems: Theoretical analysis and numerical results}.
\newblock \bibinfo{journal}{IEEE Access} \bibinfo{volume}{9}, \bibinfo{pages}{132255--132266}.
\newblock \DOIprefix\doi{10.1109/ACCESS.2021.3114969}.
\bibitem[{Tripathy et~al.(1982)Tripathy, Prasad, Malik and Hope}]{Tripathy1982}
\bibinfo{author}{Tripathy, S.}, \bibinfo{author}{Prasad, G.D.}, \bibinfo{author}{Malik, O.}, \bibinfo{author}{Hope, G.}, \bibinfo{year}{1982}.
\newblock \bibinfo{title}{Load-flow solutions for ill-conditioned power systems by a newton-like method}.
\newblock \bibinfo{journal}{IEEE Transactions on Power Apparatus and Systems} \bibinfo{volume}{PAS-101}, \bibinfo{pages}{3648--3657}.
\newblock \DOIprefix\doi{10.1109/TPAS.1982.317050}.
\bibitem[{Weidenfeller et~al.(2022)Weidenfeller, Valor, Gacon, Tornow, Bello, Woerner and Egger}]{Weidenfeller2022}
\bibinfo{author}{Weidenfeller, J.}, \bibinfo{author}{Valor, L.C.}, \bibinfo{author}{Gacon, J.}, \bibinfo{author}{Tornow, C.}, \bibinfo{author}{Bello, L.}, \bibinfo{author}{Woerner, S.}, \bibinfo{author}{Egger, D.J.}, \bibinfo{year}{2022}.
\newblock \bibinfo{title}{Scaling of the quantum approximate optimization algorithm on superconducting qubit based hardware}.
\newblock \bibinfo{journal}{Quantum} \bibinfo{volume}{6}, \bibinfo{pages}{870}.
\newblock \DOIprefix\doi{10.22331/q-2022-12-07-870}.
\bibitem[{Xiao and Li(2024)}]{Xiao2024}
\bibinfo{author}{Xiao, Q.}, \bibinfo{author}{Li, P.}, \bibinfo{year}{2024}.
\newblock \bibinfo{title}{Probabilistic power flow computation using liouville–gaussian copula and nested cubature rule}.
\newblock \bibinfo{journal}{Computers and Electrical Engineering} \bibinfo{volume}{120}, \bibinfo{pages}{109677}.
\newblock \DOIprefix\doi{10.1016/j.compeleceng.2024.109677}.
\bibitem[{Yin et~al.(2023)Yin, Tamura, Furue, Konoshima, Kanda and Watanabe}]{Yin2023}
\bibinfo{author}{Yin, F.}, \bibinfo{author}{Tamura, H.}, \bibinfo{author}{Furue, Y.}, \bibinfo{author}{Konoshima, M.}, \bibinfo{author}{Kanda, K.}, \bibinfo{author}{Watanabe, Y.}, \bibinfo{year}{2023}.
\newblock \bibinfo{title}{Extended ising machine with additional non-quadratic cost functions}.
\newblock \bibinfo{journal}{Journal of the Physical Society of Japan} \bibinfo{volume}{92}.
\newblock \DOIprefix\doi{10.7566/JPSJ.92.034002}.

\end{thebibliography}

\end{document}